# Complementarity between survival and mortality

**Byung Mook Weon**

## Abstract

Accurate demographic functions help scientists define and understand longevity. We summarize a new demographic model, the Weon model, and show the application to the demographic data for Switzerland (1876-2002). Particularly, the Weon model simply defines the maximum longevity, which is induced in nature by the mortality dynamics. In this study, we reconsider the definition of the maximum longevity and the effectiveness for longevity by the combined effect of the survival and mortality functions. The results suggest that the mortality function should be zero at the maximum longevity, since the density function is zero but the survival function is not zero. Furthermore, the effectiveness for longevity can be maximized at the characteristic life by the complementarity between the survival and mortality functions, which suggests that there may be two parts of rectangularization for longevity. The historical trends for Switzerland (1876-2002) implies that there may be a fundamental limiting force to restrict the increase of the effectiveness. As a result, it seems that the density function is essential to define and understand the mortality dynamics, the maximum longevity, the effectiveness for longevity, the paradigm of rectangularization and the historical trends of the effectiveness by the complementarity between the survival and mortality functions.

## 1. Introduction

The study of aging has traditionally been independently approached at two levels of biological organization: at the individual and sub-individual level by gerontologists interested in the physiology of human aging, and at the population level by demographers primarily interested in patterns of survival and mortality in human populations (Gohil and Joshi, 1998). Fundamental studies of the aging process have lately attracted the interest of researchers in a variety of disciplines, linking ideas and theories from such diverse fields as biochemistry to mathematics (Weitz and Fraser, 2001). The way to characterize aging is to plot the increase in mortality rate with chronological age. The mortality rate is the probability that an individual who is alive at a particular age will die during the following age interval. The mortality rate can also be represented as the fraction of the population surviving to a particular age (or the survival rate) (Weon, 2004c).

The fundamental law of population dynamics is the Gompertz law (Gompertz, 1825), in which the human mortality rate increases roughly exponentially with increasing age at senescence. The Gompertz model is most commonly employed to compare mortality rates between different populations (Penna and Stauffer, 1996). However, no mathematical model so far, including the Gompertz model, has been suggested that can perfectly approximate the development of the mortality rate over the total life span (Kowald, 1999). Particularly in modern research findings, it seems to be obvious that the mortality rate does not increase according to the Gompertz model at the highest ages (Vaupel, 1997; Robine and Vaupel, 2002), and this deviation from the Gompertz model is a great puzzle to demographers, biologists and gerontologists. There is strong evidence from many developed countries that the rate of increase in mortality decelerates at high ages. Many of the traditional mathematical models (for instance, the Gompertz, Weibull, Heligman & Pollard, Kannisto, Quadratic and Logistic models) for the mortality rate provide poor fits to empirical population data at the highest ages (Thatcher, Kannisto and Vaupel, 1998; Yi and Vaupel, 2003).

We have recently found a useful model, the Weon model, derived from the Weibull model with an age-dependent shape parameter to describe the human survival and mortality curves (Weon, 2004a, 2004b, 2004c). In this paper, we summarize the new demographic model, the Weon model, and show the application to the demographic data for Switzerland from 1876 up to 2002. Especially, we reconsider the definition of the maximum longevity, the effectiveness for longevity, the new paradigm of rectangularization for longevity and the historical trends of the effectiveness by the complementarity or the combined effect of the survival and mortality functions.

## 2. Materials and Methods

### *2.1. Demographic functions*

Mortality distributions can be effectively summarized by any one of several complementary functions (Wilmoth, 1997). Three functions are particularly useful: the density function, the survival function and the mortality function as a function of age; Let $f$ be the (*probability*) *density function* describing the distribution of life spans in a population. The *cumulative density function*, $F$, gives the probability that an individual dies before surpassing age $t$ (especially in this case, we consider that the age $t$ is a *continuous* variable). The *survival function*, $s$, gives the complementary probability ($s = 1 - F$) that an individual is still alive at age $t$. The *mortality function*, $m$, is defined as the ratio of the density and survival functions ( $m = f / s$ ). Thus, the mortality function gives the probability density at age $t$ conditional on survival to that age. The mathematical, complementary relationships among the demographic functions are as follows,

$$m = -\frac{ds}{dt} \times \frac{1}{s} \tag{1}$$

$$f = -\frac{ds}{dt} = m \times s \tag{2}$$

In life table notation, the probability of surviving to age $x$ (the age $x$ is actually a *discrete* variable), $s$, would be denoted as $l_x / l_0$. The continuous density function, $f$, would be replaced by the discrete function, $d_x$ (or $d_x / l_0$), which gives the number (or population) of life table deaths in age interval from $x$ to $x+1$. Finally, the mortality function, $m$, would be written in the life table as $m_x$ and is known to demographers and actuaries as the *force of mortality*.

### *2.2. Demographic data*

Demographic data, the period life tables (for all sexes, 1x1) and the death rate data for the years of 1876-2002 for Switzerland, were taken from the Human Mortality Database (available at http://www.mortality.org). The age dependence of the shape parameter intrinsically makes the survival and mortality functions complex and difficult for modeling. Roughly a linear expression for the shape parameter is useful for ages 0-80 (Weon, 2004a, 2004b, 2004c). For the best fits to the demographic data over the total life span; a cubic or a quartic expression is appropriate for ages 0-20, a linear or a quadratic expression is appropriate for ages 20-80 and a quadratic expression is appropriate for ages 80+ (Weon, 2004c).

## 3. Weon model

In recent a new concept, model, methodology and principle for studying human longevity in terms of demographic basis have been introduced, developed and established by Weon (2004a, 2004b, 2004c). We call the new model the "Weon model" (Weon, 2004c), which is modified from the Weibull model with an age-dependent shape parameter. This section summarizes the concept, model, methodology, generality, definition of the mathematical limit of longevity (the maximum longevity) and complementarity principle on longevity by the Weon model and shows the application to the demographic data for Switzerland.

### *3.1. Concept*

The original concept was obtained as follows: typical human survival curves show i) a rapid decrease in survival in the first few years of life and ii) a relatively steady decrease and then an abrupt decrease near death thereafter. Interestingly, the former behaviour resembles the Weibull survival function with $\beta < 1$ and the latter behaviour seems to follow the case of $\beta >> 1$. With this in mind, it could be assumed that shape parameter is a function of age (Weon, 2004b). The new model is completely different from the Weibull model in terms of the 'age dependence of the shape parameter'. It is especially noted that the shape parameter can indicate a 'rectangularity' of the survival curve. The reason for this is that as the value of the shape parameter becomes a high value, the shape of the survival curve approaches a further rectangular shape (Weon, 2004a).

### *3.2. Model*

The Weon model is derived from the Weibull survival function and it is simply described by two parameters, the age-dependent shape parameter and the characteristic life. The age-dependent shape parameter enables us to model the survival and mortality functions and it is expressed as follows,

$$s = \exp(-(t/\alpha)^{\beta}) \tag{3}$$

$$m = (t/\alpha)^{\beta} \times [\frac{\beta}{t} + \ln(t/\alpha) \times \frac{d\beta}{dt}] \tag{4}$$

where $\alpha$ denotes the *characteristic life* (or the scale parameter, $t = \alpha$ when $s = \exp(-1)$) and $\beta$ denotes the *shape parameter* as a "function of age". The Weon model is completely different with the Weibull model in the age dependence of the shape parameter. The fact that the shape parameter for humans is a function of age is valid with a certain degree of universality in many countries (Weon, 2004a, 2004b, 2004c). The density function by the Weon model can be expressed the multiplication of the equation (3) and (4) by the mathematical relationship of '$f = m \times s$'.

### *3.3. Methodology*

We could evaluate the age dependence of the shape parameter to determine an adequate mathematical expression of the shape parameter, after determination of the characteristic life graphically in the survival curve. Conveniently, the value of the characteristic life is always found at the duration for survival to be '$\exp(-1)$', this is known as the characteristic life. This feature gives the advantage of looking for the value of $\alpha$ simply by graphical analysis of the survival curve. In turn, with the observed value of $\alpha$, we can plot the shape parameter as a function of age by the mathematical equivalence of '$\beta = \ln(-\ln s)/\ln(t/\alpha)$'. If $\beta$ is not constant with age, this obviously implies that '$\beta$ is a function of age'. For example, we can see the age dependence of the shape parameter for Switzerland (2002) in Fig. 1 when $\alpha = 86.98$ years.

In empirical practice, we could successfully use a polynomial expression for modeling the shape parameter as a function of age: $\beta = \beta_0 + \beta_1 t + \beta_2 t^2 + \ldots$, where the associated coefficients could be determined by a regression analysis in the plot of shape parameter curve. And thus, the derivative of $\beta$ is obtained as follows: $d\beta/dt = \beta_1 + 2\beta_2 t + \ldots$, which indicates again that the shape parameter for humans is a function of age. Roughly a linear expression is useful for ages 0-80. But for the best fits to the demographic data over the total life span; a cubic or a quartic expression is appropriate for ages 0-20, a linear or a quadratic expression is appropriate for ages 20-80 and a quadratic expression is appropriate for ages 80+ (Weon, 2004c).

On the other hand, $\beta$ mathematically approaches infinity as the age $t$ approaches the value of $\alpha$ or the denominator '$\ln(t/\alpha)$' approaches zero. This feature of $\beta$ can leave 'trace of $\alpha$' in the plot of $\beta$, so we can observe variations of $\beta$ and $\alpha$ at once in the plot of the shape parameters. If $\beta$ (except for the mathematical singularity or trace of $\alpha$) can be expressed by an adequate mathematical function, the survival and mortality functions can be calculated by the mathematically expressed $\beta$. Only two parameters, $\alpha$ and $\beta$, determine the survival and mortality functions. In the case of Switzerland (2002), when the characteristic life is evaluated to be 86.98 years, the trace of $\alpha$ can be observed at the age 87 near the characteristic life in Fig. 1.

### *3.4. Generality*

The Gompertz model (Gompertz, 1825) and the Weibull model (Weibull, 1951) are the most generally used models at present (Gavrilov and Gavrilova, 2001). Interestingly, the Gompertz model is more commonly used to describe biological systems, whereas the Weibull model is more commonly applicable to technical devices (Gavrilov and Gavrilova, 2001). In the previous paper (Weon, 2004c), we could see that the traditional models, the Gompertz and Weibull models, may be generalized by the Weon model on the basis of the fact that shape parameter is a function of age by the approximate relationship of '$\ln m \propto \beta$' after adulthood (for ages ~20+) (Weon, 2004c). The Weon model approximates the Gompertz model when '$\beta \propto t$' and the Weibull model when '$\beta = constant$'. We could see that the Gompertz model is a special case of a linear expression for $\beta$ and the Weibull model is a special case of a constant shape parameter. Particularly, the mortality rate would deviate from the Gompertz model when $\beta$ shows a non-linear behavior (before age ~20 or after age 80). Thus, $\beta$ is a measure of the deviation from the Gompertz model (Weon, 2004c).

Particularly for aging patterns, it is the age dependence of the shape parameter that distinguishes humans from technical devices. It seems to show the difference between humans and technical devices in terms of '*robustness*'. The fundamental difference for robustness between biological systems and technical devices is obvious (Gavrilov and Gavrilova, 2001). In the previous papers (Weon, 2004a, 2004b, 2004c), the age-dependent shape parameter is changed from approximately 0.5 to 10 with age for the typical demographic curves. This feature is in great contrast to technical devices typically having a constant shape parameter (Nelson, 1990). We attribute the age dependence of the shape parameter to the resistance to aging, or the nature (the *homeostasis* and the *adaptation*) of biological systems to maintain stability and to survive (Weon, 2004a). A fundamental principle related to the age-dependent shape parameter for humans was suggested in the recent paper (Weon, 2004c), which will be explained in the section 3.6.

### *3.5. Mathematical limit of longevity (maximum longevity)*

In general, the mortality rate should be mathematically positive ( $m > 0$ ). Therefore, the criterion for the mathematical limit of longevity, implying the maximum longevity which is able to be determined by the mortality dynamics in nature, can be given by (Weon, 2004c),

$$\frac{d\beta}{dt} > -\frac{\beta}{t\ln(t/\alpha)} \qquad (5)$$

We successfully used a quadratic expression for the description of the shape parameter after age 80. For example, see the case of Switzerland (2002) from age 80 to 109 in Fig. 2. Interestingly, the quadratic coefficient ( $\beta_2$ ) is important to evaluate the mathematical limit of longevity, since it determines the slope with age in the derivative ( $\beta_1 + 2\beta_2 t$ ) of the quadratic expression of the shape parameter (Weon, 2004c).

Specifically, the mortality curves for higher ages (110+) are important to understand the human longevity. According to the Weon model, the quadratic expression for ages 80-109 is valid with a certain degree of university in many modern developed countries, which enables us to predict that the mortality rate decreases after a plateau around ages 110-115 and the mathematical limit of longevity emerges around ages 120-130 (Weon, 2004c). If the quadratic expression is valid for ages 110+, we are able to predict the mortality rate at the highest ages. The pattern of the mortality dynamics (deceleration, plateau and decrease) at the highest ages by the Weon model is consistent with the other assertions (for instance, Vaupel et al., 1998; Robine and Vaupel, 2002; Helfand and Inouye, 2002).

### *3.6. Complementarity principle on longevity*

In recent we suggested a fundamental principle on longevity (Weon, 2004c). We wish to explain it in brief. The essence of the Weon model is the age dependence of the shape parameter. What is the origin of the age-dependent shape parameter? According to the Weon model, in principle for the highest value of $s$ or for longevity at all times, the shape parameter should be variable according to the characteristic life; "for longevity, $\beta$ increases at $t < \alpha$ but it decreases at $t > \alpha$." See an example of the longevity tendency in Fig. 3. This is attributable to the nature of biological systems to strive to survive healthier and longer. According to the Weon model, empirically the quadratic coefficient ($\beta_2$) indicates the decrease of $\beta$ at $t > \alpha$. Thus, the longevity tends to increase with increasing $\beta_2$. On the other hand, we could see that the mortality dynamics (deceleration, plateau and decrease) are a consequence of the decrease of the shape parameter at $t > \alpha$. According to the Weon model, the quadratic expression is obviously related with the mortality dynamics at $t > \alpha$, which induces the mathematical limit (the mortality rate to be mathematically zero, implying the maximum longevity). Interestingly, the mathematical limit tends to decrease with increasing quadratic coefficient ( $\beta_2$ ). It seems that the mathematical limit decreases as the longevity increases, which shows "complementary" aspects. It is very interesting that the reason for longevity, especially in terms of the decrease of the shape parameter for ages after characteristic life, may be the reason for limit of longevity in nature (Weon, 2004c).

## 4. Results and Discussion

In this study, it seems to be obvious that the mortality function ( $m$ ) is susceptible to the density function ( $f = -ds/dt$ ) before age ~60 but it is sensitive to the reciprocal of the survival function ( $1/s$ ) thereafter. However, the density function ( $f = -ds/dt$ ) seems to drive again the mortality function ( $m$ ) to decelerate at the highest ages. See an example for Switzerland (2002) in Fig. 4. According to the Weon model, the quadratic expression for the shape parameter at ages 80-109 results in that the mortality function is zero at the maximum longevity by the mortality dynamics in nature (Weon, 2004c). It seems that the density function is essential to understand the mortality dynamics. Especially in this study, we reconsider the definition for the maximum longevity and the demographic meaning of the density function by the complementarity between the survival and mortality functions. The demographic analysis for Switzerland (1876-2002) supports the findings by the Weon model. In the following sections, we think about the definition of maximum longevity by the Weon, the effectiveness for longevity by the complementarity between the survival and mortality functions, the new paradigm of rectangularization for longevity by the effectiveness and the historical trends of the effectiveness for Switzerland (1876-2002).

### *4.1. Definition of maximum longevity*

In general, the term of "*longevity*" means the "*duration of life*". In a sense, the "*maximum longevity*" can be used to mean the "*maximum duration of life*" of a given population. However, what we know is the "*maximum age at death*", which means the oldest age at death observed in a given population during a given time period (Vallin and Meslé, 2001). Perhaps the most common notion of a limit in the study of human longevity is the *limited-life-span hypothesis*, which states that there exists some age $\omega$ beyond which there can be no survivors. This hypothesis can be expressed by any one of the following three formulas (Wilmoth, 1997): "$f = 0$, $s = 0$ or $\lim_{t \to \omega} m = \infty$ ($t \geq \omega$)." By the way, according to the Weon model, the survival function is not zero, although it has extremely low values at the highest ages, but that the mortality function can be zero at the maximum longevity. See an example for Switzerland in Fig. 5. The survival curves for 1876 and 2002 are not zero when they are extrapolated by the estimation of the shape parameters for ages 80-109. Therefore, the Weon model suggests that the maximum longevity can be defined as follows: "at $t = \omega$, $f = 0$ and $m = 0$, because of $s \neq 0$."

If that is the case, why the mortality rate approaches zero at the maximum longevity in the Weon model? We think about what will happen to the survival rate at the maximum longevity. In fact the survival rate approaches zero, but it is not zero. The survival function is a kind of cumulative probability function as the complementary function of the cumulative density function ($s = 1 - F$), indicating the probability of survival beyond age $t$. Even at the maximum longevity ($\omega$), the survival function need not be always zero. Instead, the decrease rate of the survival function with age ($-ds/dt$; the minus indicates the decrease) should be zero at the maximum longevity. This means that the density function ($f = -ds/dt$) should be zero and thus the mortality function ($m = f/s$) should be zero at the maximum longevity, since the survival function is not zero ($s \neq 0$). Consequently, the maximum longevity can be simply defined by the following simple mathematical expression,

$$-\frac{ds}{dt} = 0 \tag{6}$$

This simple mathematical expression for the maximum longevity makes sense and comprehends the definitions by the density and mortality functions as follows: "at $t = \omega$, $f = 0$ and $m = 0$, because of $s \neq 0$." The values for the maximum longevity calculated from the equation (5) and (6) are mathematically identical.

We confirm the fact that the mortality function should be zero at the maximum longevity, since the density function is zero but the survival function is not zero at the moment. The functions beyond the maximum longevity have no reality. See an example for modeling the mortality function for Switzerland (2002) in Fig. 6. The mortality function extrapolated by the Weon model approaches zero at the maximum longevity, which is due to the nature of the density function: That is, the decrease rate of the survival function with age ($-ds/dt$) or the density function should be zero at the maximum longevity.

### *4.2. Effectiveness for longevity*

It seems that the density function indicates the *effectiveness* for longevity between the survival and mortality functions, $f = m \times s$, since the mortality rate tends to increase with decreasing the survival rate, which is due to the *complementarity* between the survival and mortality functions; “For longevity, individuals tend to reduce the mortality rate but strive to improve the survival rate.” It is expected that there may exist the ‘maximum’ *effectiveness* between the survival and mortality rates at the ‘characteristic life’. See an example for Switzerland (2002) in Fig. 7. The effectiveness for longevity is maximized at the characteristic life ($\alpha$ ~87 years) and it is zero at the maximum longevity ($\omega$ ~124 years) for Switzerland (2002).

### *4.3. Two-part rectangularization*

If that is the case, it can be suggested that since the most effective combination between the survival and mortality functions for longevity occurs at the characteristic life, there may be two parts of rectangularization for longevity as shown in Fig. 8 as follows: The survival function is rectangularized by the *increase* of the shape parameter before characteristic life ($0 \sim \alpha$) – it is the *first part*. The survival function is rectangularized by the *decrease* of the shape parameter after characteristic life ($\alpha \sim \omega$) – it is the *second part*. The first and second parts can be overlapped as one rectangularization as the characteristic life approaches the maximum longevity ($\alpha \rightarrow \omega$). This paradigm of rectangularization for longevity makes sense and comprehends the conventional paradigm of rectangularization (Fried, 1980; Eakin and Witten, 1995).

*4.4. Historical trends*

The historical trends of the density function for Switzerland from 1876 up to 2002 are seen in Fig. 9. It tells the trends of the *effectiveness* for longevity by increasing the survival rate and decreasing the mortality rate over time. The density function shifts to the right and upward direction, which indicates that the most effective combination between the survival and mortality functions for longevity *effectively* increases as the characteristic life increases. However, it seems that the exponential decrease rates of the density function after the characteristic life does not significantly increases over time. This implies that a fundamental limiting force may restrict the *effectiveness* for longevity over time. It is attributed to the *complementarity principle* on longevity; "for longevity, $\beta$ increases at $t < \alpha$ but it decreases at $t > \alpha$, resulting in that the mathematical limit (the maximum longevity) decreases as the longevity increases in nature (Weon, 2004c)".

**5. Conclusions**

In this paper, we summarize the new demographic model, the Weon model, and show the application to the demographic data for Switzerland (1876-2002). In this study, it seems to be obvious that the density function is essential to understand the mortality dynamics. We reconsider the definition of maximum longevity, the effectiveness for longevity by the complementarity between the survival and mortality functions, the new paradigm of rectangularization for longevity by the effectiveness and the historical trends of the effectiveness for Switzerland (1876-2002). The results suggest that the mortality function should be zero at the maximum longevity, since the density function is zero but the survival function is not zero. Furthermore, the effectiveness for longevity can be maximized at the characteristic life by the complementarity between the survival and mortality functions. Since the most effective combination between the survival and mortality functions for longevity occurs at the characteristic life, there may be two parts of rectangularization for longevity according to the characteristic life. The historical trends for Switzerland (1876-2002) tells that the most effective combination between the survival and mortality functions for longevity effectively increases as the characteristic life increases over time, however the exponential decrease rates of the density function after the characteristic life does not significantly increases, which implies that a fundamental limiting force may restrict the effectiveness for longevity. As a result, it seems that the density function is essential to define and understand the mortality dynamics, the maximum longevity, the effectiveness for longevity, the paradigm of rectangularization and the historical trends of the effectiveness by the complementarity between the survival and mortality functions.

**References**

Eakin, T., Witten, M., 1995. How square is the survival-curve of a given species. Exp. Gerontol. 30, 33-64.

Fried, J.F., 1980. Aging, natural death, and the compression of mortality. N. Engl. J. Med. 303, 130-135.

Gavrilov, L.A., Gavrilova, N.S., 2001. The reliability theory of aging and longevity. J. Theor. Biol. 213, 527-545.

Gohil, V., Joshi, A., 1998. Modelling the evolution of rates of ageing. Resonance 3, 67-72.

Gompertz, B., 1825. On the nature of the function expressive of the law of human mortality and on a new mode of determining life contingencies. Philos. Trans. Roy. Soc. London Ser. A 115, 513-585.

Helfand, S.L., Inouye, S.K., 2002. Rejuvenating views of the ageing process. Nature Reviews Genetics 3, 149-153.

Human Mortality Database. The University of California, Berkeley (USA), and The Max Planck Institute for Demographic Research (Germany). Available at: http://www.mortality.org (The data was downloaded on 17 February 2004).

Kowald, A., 1999. Theoretical Gompertzian implications on life span variability among genotypically identical animals. Mech. Ageing Dev. 110, 101-107.

Nelson, W., 1990. Accelerated testing: statistical functions, test plans, and data analysis. New York: Wiley: pp. 63-65.

Oeppen, J., Vaupel, J.W., 2002. Broken limits to life expectancy. Science 296, 1029-1031.

Penna, T.J.P., Stauffer, D., 1996. Bit-string aging model and German population. Z. Phys. B 101, 469-470.

Robine, J.M., Vaupel, J.W., 2002. Emergence of supercentenarians in low mortality countries. North American Actuarial Journal 6, 54-63.

Thatcher, A.R., Kannisto, V., Vaupel, J.W., 1998. The force of mortality at ages 80 to 120. Odense Monographs on Population Aging, Vol 5., Odense, Odense University Press. (http://www.demogr.mpg.de/Papers/Books/Monograph5/ForMort.htm).

Vallin, J., Meslé, F., 2001. Living beyond the age of 1000. Population and Sociétés. 365, 1-4.

Vaupel, J.W., 1997. Trajectories of mortality at advanced ages. In: Wachter, K.W., Finch, C.E., (Ed.), Between zeus and the salmon: the biodemography of longevity, Washington DC, National Academic Press, pp. 17-37.

Vaupel, J.W., Carey, J.R., Christensen, K., Johnson, T.E., Yashin, A.I., Holm, N.V., Iachine, I.A., Khazaeli, A.A., Liedo, P., Longo, V.D., Zeng, Y., Manton, K.G., Curtsinger, J.W., 1998. Biodemographic trajectories of longevity. Science 280, 855-860.

Weibull, W., 1951. A statistical distribution function of wide applicability. J. Appl. Mech. 18, 293-297.

Weitz, J.S., Fraser, H.B., 2001. Explaining mortality rate plateau. Proc. Natl. Acad. Sci. USA 98, 15383-15386.

Weon, B.M., 2004a. Analysis of trends in human longevity by new model. Submitted to Demographic Research. The pre-print is available at: http://arxiv.org/abs/q-bio/0402011.

Weon, B.M., 2004b. General functions for human survival and mortality. Submitted to Mech. Ageing Dev. The pre-print is available at: http://arxiv.org/abs/q-bio/0402013.

Weon, B.M., 2004c. Complementarity principle on human longevity. Submitted to Demographic Research. The pre-print is available at: http://arxiv.org/abs/q-bio/0403010.

Wilmoth, J.R., 1997. Trajectories of mortality at advanced ages. In: Wachter, K.W., Finch, C.E., (Ed.), Between zeus and the salmon: the biodemography of longevity, Washington DC, National Academic Press, pp. 38-64.

Wilmoth, J.R., Deegan, L.J., Lundstöm, H., Horiuchi, S., 2000. Increase of maximum life-span in Sweden, 1861-1999. Science 289, 2366-2368.

Yi, Z., Vaupel, J.W., 2003. Oldest-old mortality in China. Demographic Research [Online] 8. Available at: http://www.demographic-research.org/volumes/vol8/7.

**Acknowledgements**

The author thanks to the Human Mortality Database (Dr. John R. Wilmoth, as a director, in The University of California, Berkeley and Dr. Vladimir Shkolnikov, as a co-director, in The Max Planck Institute for Demographic Research) for allowing anyone to access the demographic data for research.

**Figures**

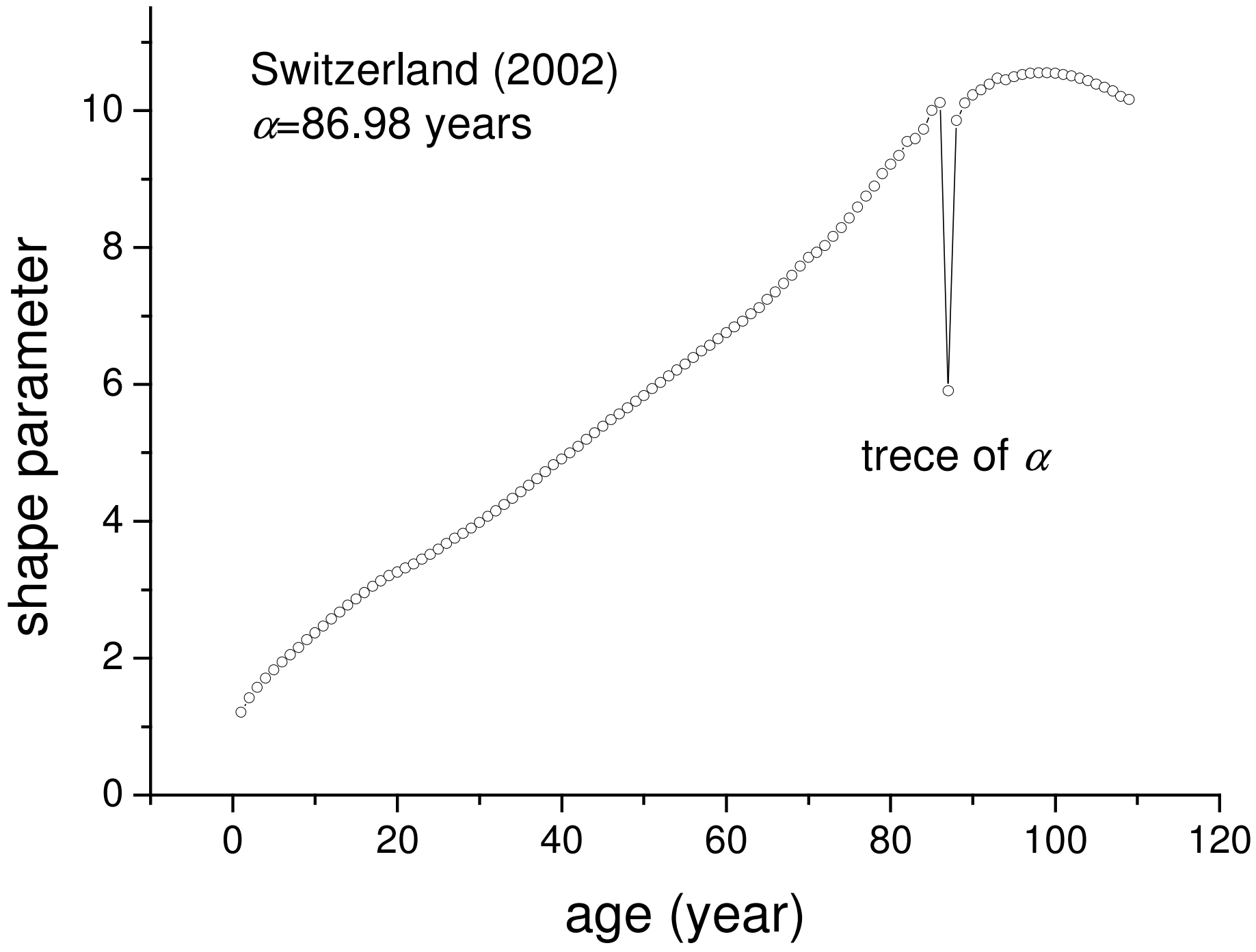

Fig. 1. The shape parameter as a function of age for Switzerland (2002) when the characteristic life is 86.98 years.

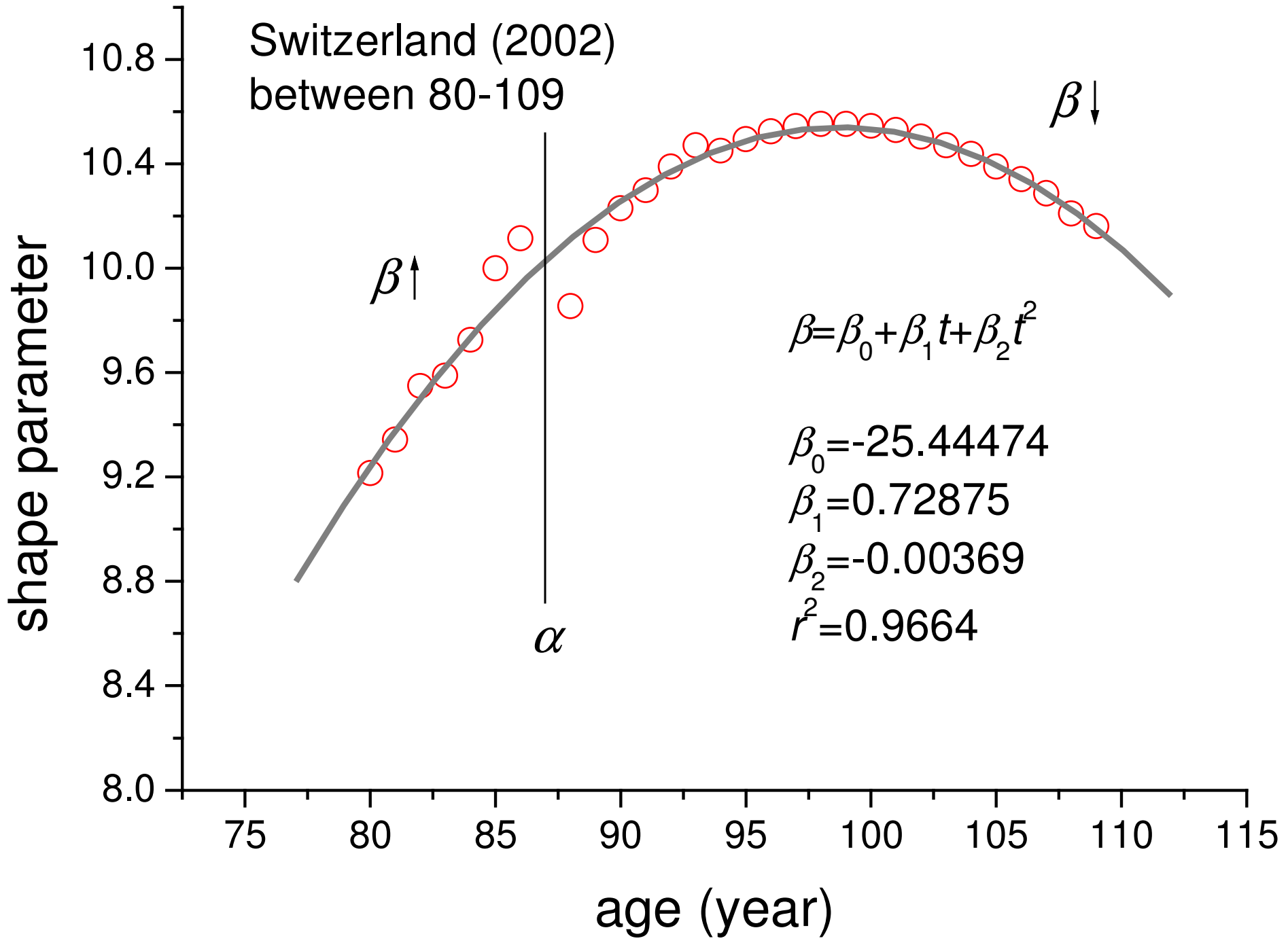

Fig. 2. Modeling the shape parameter as a quadratic expression between ages 80-109 for Switzerland (2002).

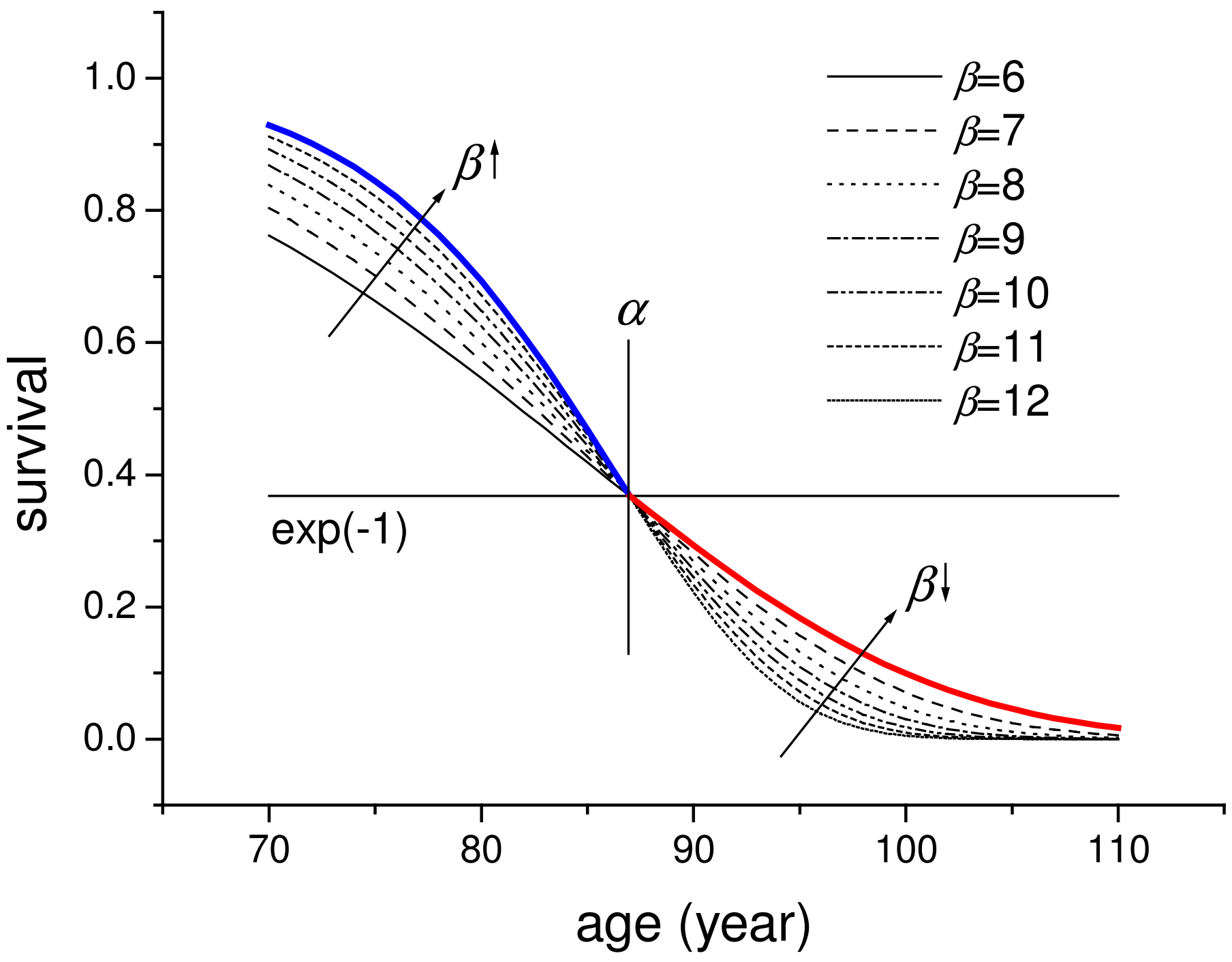

Fig. 3. Tendency for longevity by adapting the shape parameter around the characteristic life.

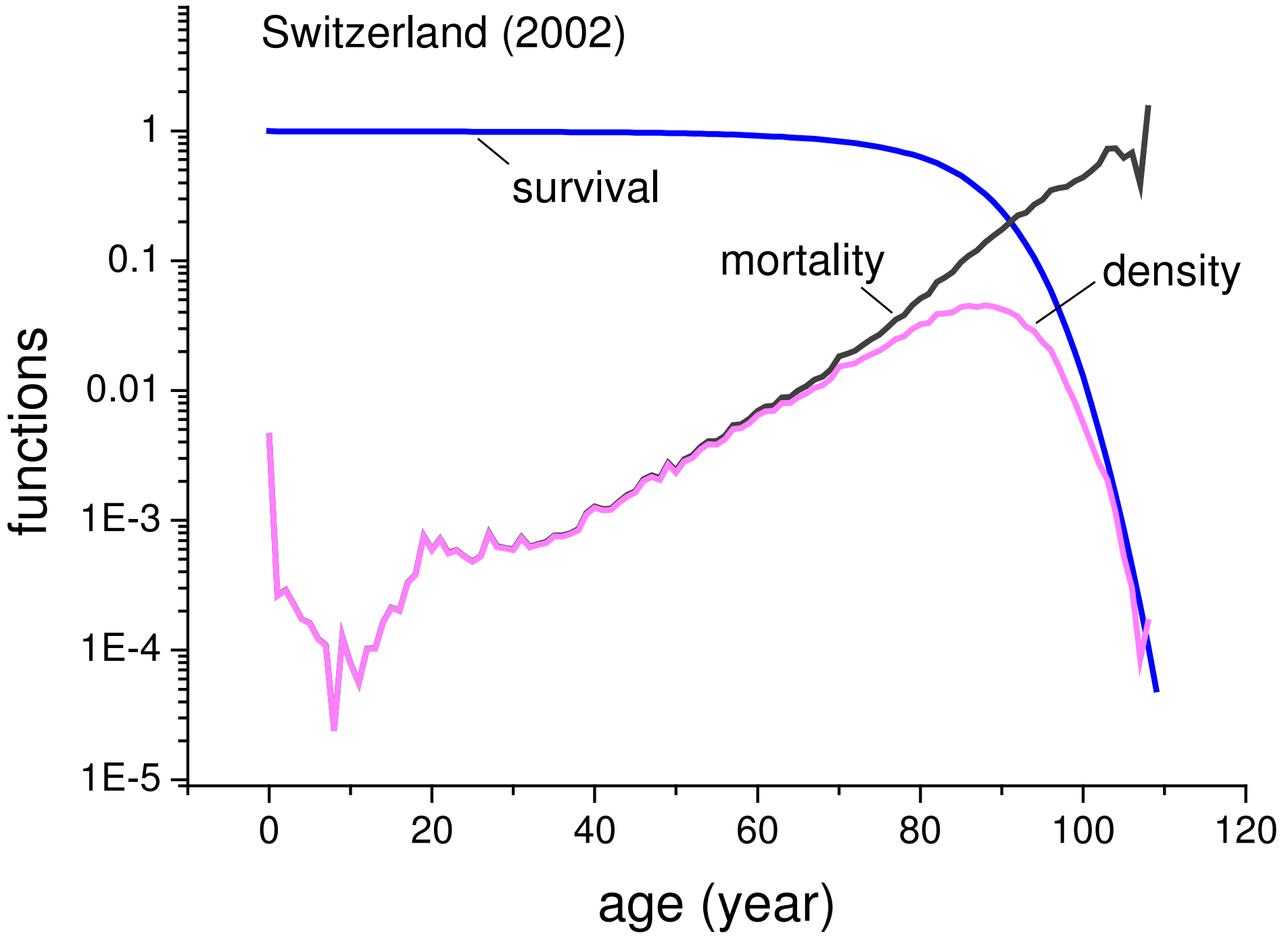

Fig. 4. Complementarity of the demographic functions, the survival, mortality and density functions as a function of age for Switzerland (2002).

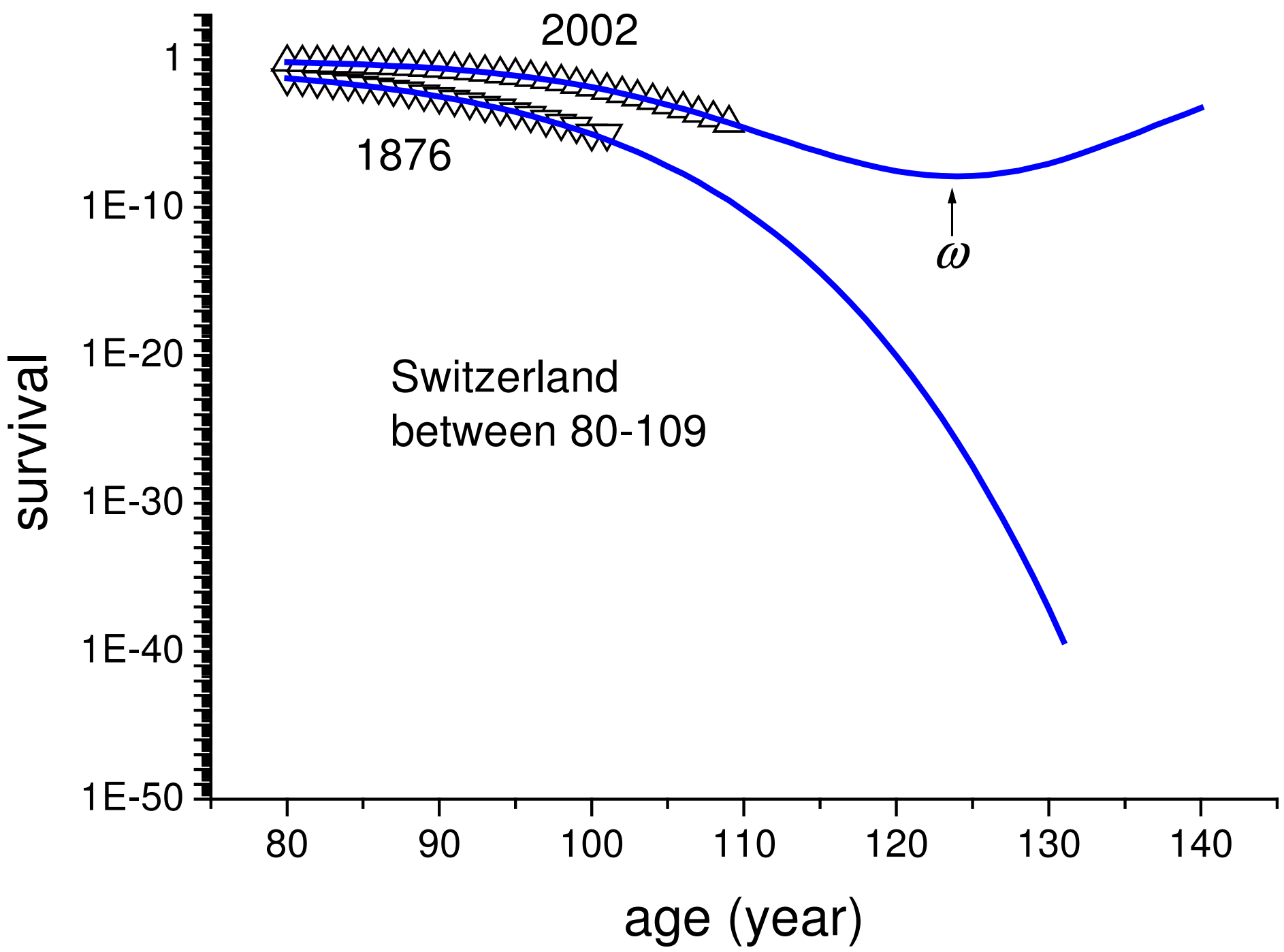

Fig. 5. Comparison between the data of the survival rate and the model of the survival function through modeling the shape parameter for Switzerland (1876 and 2002).

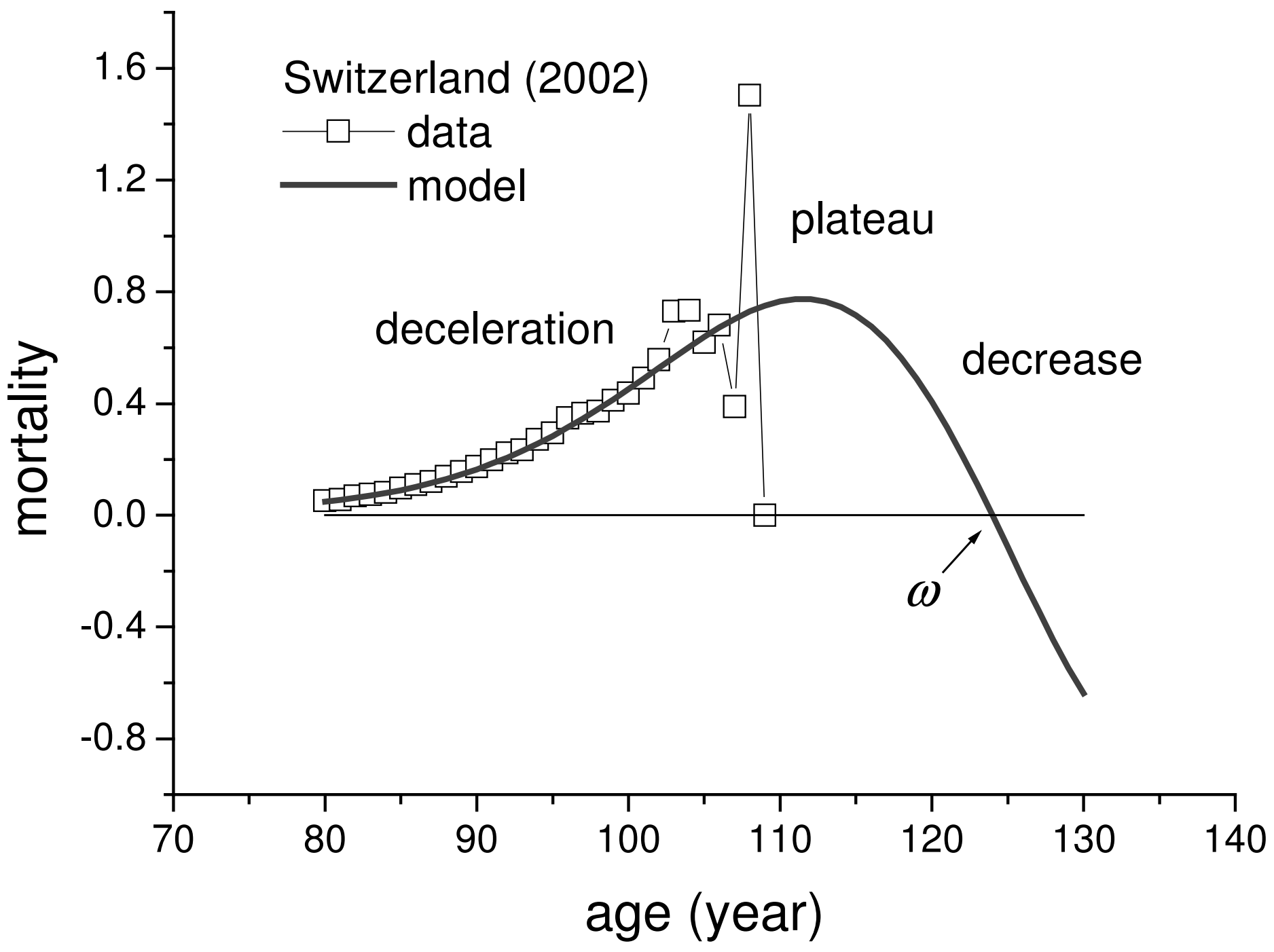

Fig. 6. Comparison between the data of the mortality rate and the model of the mortality function through modeling the shape parameter for Switzerland (2002).

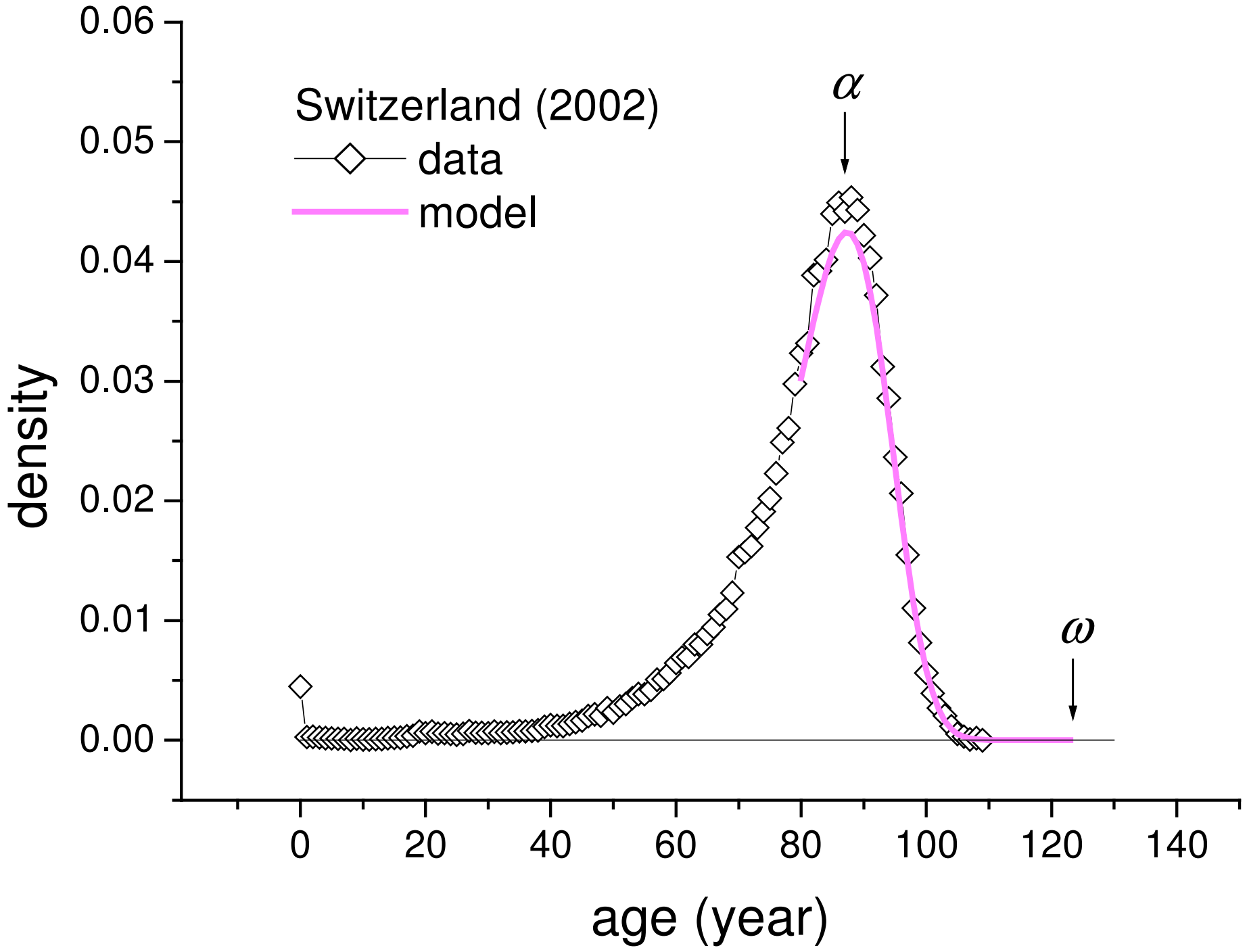

Fig. 7. Comparison between the data of the density function and the model through modeling the survival and mortality functions for Switzerland (2002).

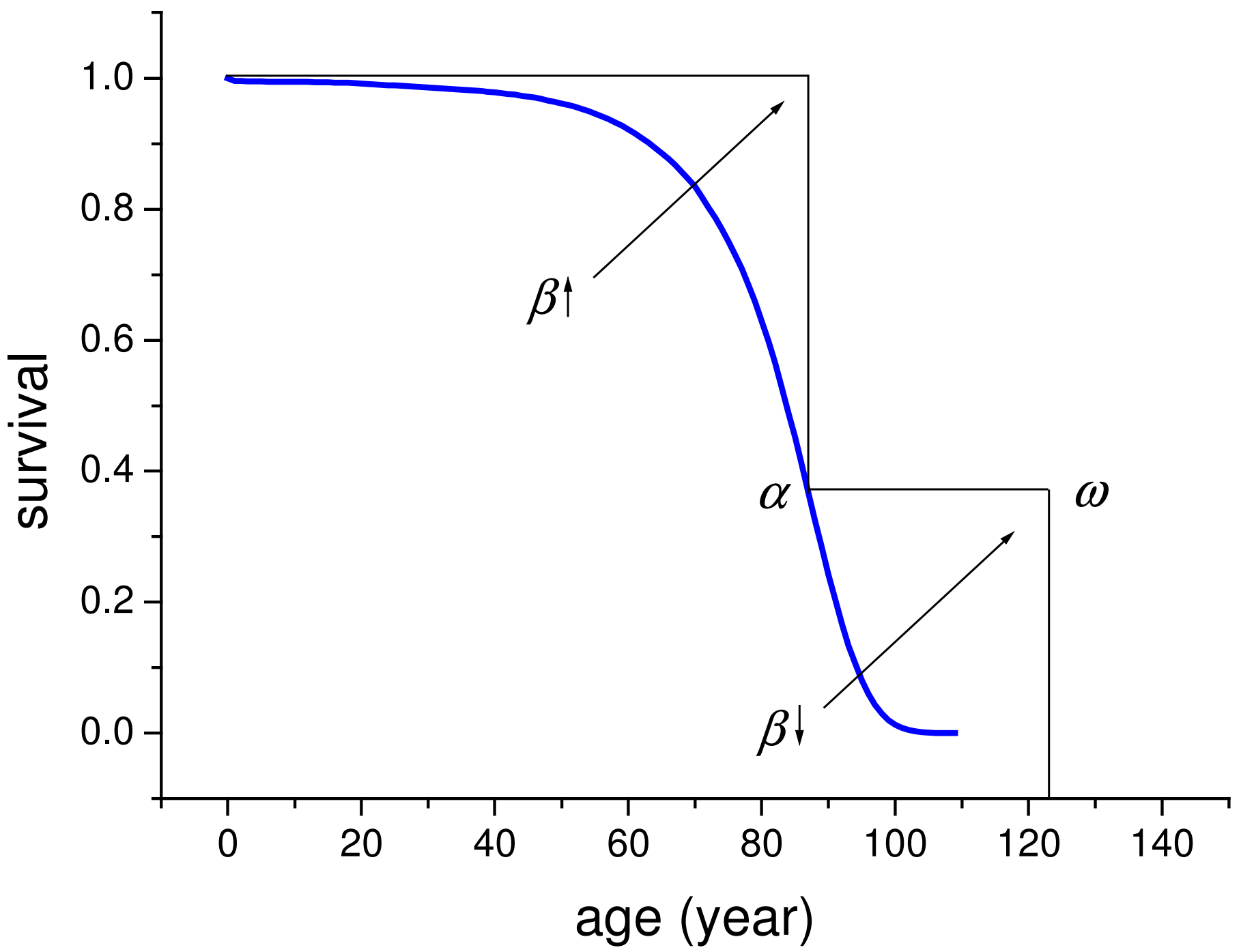

Fig. 8. Demonstration of two parts of rectangularization for longevity.

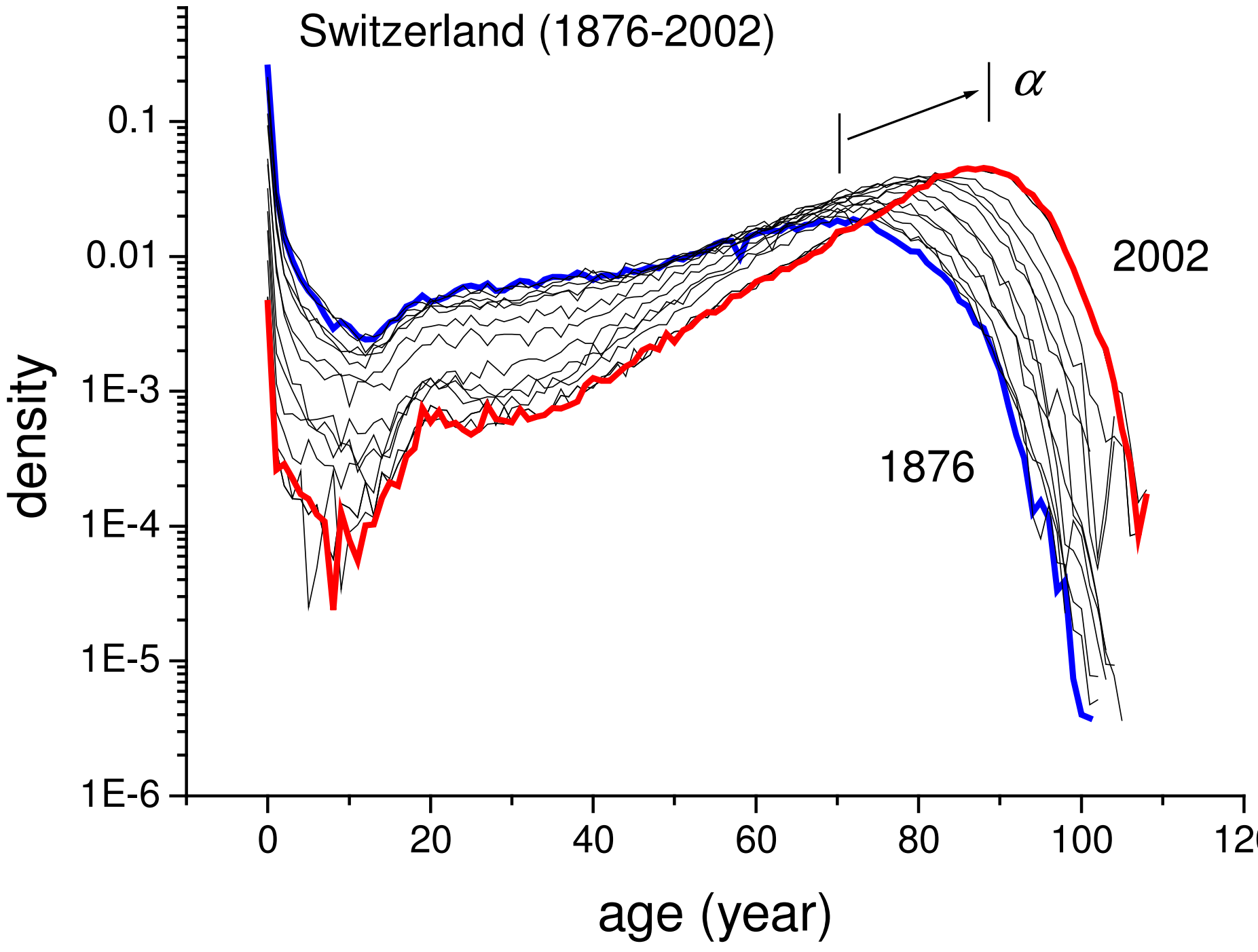

Fig. 9. Historical trends of the effectiveness or the density function for Switzerland.